# Structural evolution of the elastic properties in nano-structured AlN films

R. J. Jiménez Riobóo[1,a], V. Brien[2], P. Pigeat[2]

1. Instituto de Ciencia de Materiales de Madrid (CSIC), Campus de Cantoblanco s/n E-28049-Madrid, Spain
2. CNRS/Nancy-Université, (LPMIA, UMR CNRS 7040), Boulevard des Aiguillettes, B.P. 239, F-54506 Vandœuvre-lès-Nancy, France

a) Author to whom correspondence should be addressed. Electronic address: rjimenez@icmm.csic.es

A study of the transverse acoustic phonons on nano-structured AlN films has been carried out by using high-resolution micro-Brillouin spectroscopy. Dense film shave been deposited by radio frequency (r.f.) magnetron sputtering under ultra-high vacuum at room temperature. Films with different morphologies were prepared and investigated by transmission electron microscopy and Brillouin Spectroscopy (equiaxed nano-sized, nanocolumnar grains and amorphous phase). Results show a dependence of the transverse modes on the nano-structure. The nano-columnar film exhibits two transversal modes as expected for the AlN würtzite while the equiaxed nanosized and the amorphous films only exhibit one isotropic transverse mode as expected in amorphous materials. One important result is that the sound propagation velocity in the AlN amorphous phase is higher than the one in the non-textured nano-crystalline phase. This phenomenon has, however, already been observed in ferroelectric ceramics.

## I. INTRODUCTION

Due to an outstanding panel of interesting properties AlN offers opportunities to electronic and photonic applications in the field of sensors (optics, mechanical field…) or in micro-systems (adaptive optics, optical switch, pump…) [1]. Recently, the direct band gap semi-conducting compound, aluminium nitride has also appeared as a good candidate to host rare earth ions (notably erbium) and enhance their photoluminescence [2–4] promising thus interesting applications in the telecommunication domain. Although the physical mechanisms of electrons–photons interactions explaining the photoluminescence in doped semiconductors are not fully understood yet, some hypotheses are, however, mentioned in some papers [5–8]. First, it appears that the size and the structure of the band gap of the host material is a key parameter. On the one hand, the matrix defects and their energy levels are mentioned as playing an important role. On the other hand, the confinement of the material is known to modify the gap and these defects. Therefore, an optimisation work on the process (to get specific shape or size of grains) has to be made to find ideal nano-structures so that an optimal structural configuration promotes the efficiency of the radiative transfers between the host material and the dopant [9]. In other respects, it seems that the efficiency of luminescence in doped nano-structured materials can be deteriorated by energy transfer to phonons leading to non-radiative relaxation mechanisms [10].

To make progress in the understanding of the mechanisms of photoluminescence, it is therefore very interesting to characterise precisely the phonons inside the matrix before doping the films. This characterisation is all the more necessary and crucial as other physical responses (particularly mechanical) of crystallised materials are now known to exhibit anomalies as the grain size is reduced down to the nano-meter scale [11] and all the more so as bond order deficiencies in nano-structures have strong impacts on acoustics and photonics [12]. For instance, some other piezoelectric materials ($PbTiO_3$, $BaTiO_3$, $Si_3N_4$) are known to undergo crystallite-size-driven phase/structure transitions and hence modification of their properties [13–17] when the sizes of the powder grains come around a few tens of nano-meters. This could be due to the presence of dangling polar bonds forming thus localised defect and electrical polar moments [14]. Nano-morphologies and their characteristics (shape and size) could then strongly modify the phonon modes (acoustic and optical). Brillouin and Raman spectroscopies both result from photon–phonon interactions, and are thus particularly well adapted techniques to characterise the different phonon modes. As Raman experiments are currently in progress, this article presents the measurements obtained by High Resolution micro-Brillouin Spectroscopy. Three kinds of AlN morphologies were synthesised and analysed: amorphous (a*), granular with equiaxed crystallites (nc) and granular with nano-columns (ncol). The results are presented as a function of the nano-structure in order to highlight the structure effects on the phonon dynamics.

## II. EXPERIMENTAL
### A. SYNTHESIS

In the prospect of making a material thanks to a technology easily transferable to industry the authors have used radio frequency (r.f.) magnetron sputtering to prepare the films. The different nano-structures were prepared thanks to adequately chosen different process parameters (details on process can be found in previous works) [18, 19]. In this study, AlN thin films were sputtered on [001] mono-crystalline silicon substrates. Radio frequency magnetron sputtering was performed under ultra-high vacuum (UHV) by using an aluminium target with the purity of 99.999% in a mixed Ar/$N_2$ atmosphere. The gas purities were 99.999 %. The sputtering system was pumped down to a residual pressure of $1.10^{-6}$ Pa of $H_2O$ controlled by mass spectroscopy. In this study, bias voltage is set to 0 Volt. The r.f. power (W) delivered by the 13.56 MHz generator was in the range 50 W – 300 W. The variable parameters used to get different morphologies in the films are P, W and the quality of the gas regarding its content in oxygen. The adequate elaboration parameters chosen for preparing the samples are compiled in Table I. The thickness of the samples was controlled in real-time during the growth of the layer thanks to an interferential optical reflectometer and was confirmed by transmission electron microscopy (TEM) cross-section observations [20]. This last characterisation also showed that the films contain no porosity and that they are fully dense. Other previous optical characterisation confirmed this observation [21].

### B. TECHNIQUES OF CHARACTERISATION

Transmission electron microscopy (TEM) observations were performed on a PHILIPS CM20 microscope operating at an accelerating voltage of 200 kV using the technique of microcleavage of samples. Both cross sections and top views of the films were taken. Only cross sections images are presented.

In order to study the influence of the nano-morphology of AlN films on the acoustic propagation modes, and due to the small thickness of the samples, instead of the classical High Resolution Brillouin Spectroscopy, high-resolution micro-Brillouin spectroscopy (HRmBS) performed at room temperature was the experimental technique chosen. In the special case of mBS, the usual optical system was substituted for an Olympus BX51 microscope with a 20x objective. At this magnification, the spot size is equal to 4 μm diameter. The experimental setup of the BS technique was already described elsewhere [22, 23]. It can be summarized as follows: the light source was a 2060 Beamlok® Spectra Physics Ar+ ion laser provided with an intra-cavity temperature stabilized single-mode and single-frequency z-lok® etalon ($\lambda_0$ = 514.5nm). The scattered light was analysed using a Sandercock-type 3+3 tandem Fabry-Pérot interferometer [24]. The typical values for finesse and contrast were 150 and $10^9$, respectively. The use of a microscope to perform the HRmBS technique implies the backscattering geometry as the only one suitable, and the samples must be tilted (incident beam non perpendicular to the surface) in order to avoid the direct reflection in the photo-multiplier. The tilt angle (sagittal angle α) was set to 55°. As the AlN samples are transparent films deposited on reflecting substrates, there is a additional scattering geometry to the classical backscattering one that can be observed. This scattering geometry (noted as 2αA) defines the phonon wave-vector within the sample plane [25] and its corresponding wavelength is: $\Lambda = \frac{\lambda_0}{2(\sin\alpha)}$, with $\lambda_0$ being the laser wavelength in vacuum. The propagation velocity of the related acoustic waves is v = f x Λ with f being the Brillouin frequency shift observed in this scattering geometry.

### III. RESULTS

Figure 1 shows the (bright field or dark field) TEM images and the electron diffraction pattern recorded for each of the three nano-structures. Two samples are crystallized and one is amorphous. These first two samples A and B indeed present (Figs. 1a and b) the characteristic peaks of the AlN hexagonal würtzite-type structure (J.C.P.D.S file no. 25-1133). The columns grew on
an amorphous adaptation layer of a few nano-meters [20, 26]. Sample A is then called ''ncol'' after its nano-structure. Sample B (Fig. 1b) is made of equiaxed grains. Their average size is 3 nm. Sample B is then called ''nc''. Sample C is not a crystallised film (Fig. 1c). It is amorphous as the TEM image shows it: the electron image and the electron diffraction pattern are typical of amorphous structures [27]. The image exhibits an ''orange skin'' contrast and the electronic pattern is made of two diffuse rings [27]. Sample C is amorphous and it is renamed ''a*''.

Figures 2, 3 and 4 show the obtained HRmBS spectra (the frequency range containing the elastic scattering line or central Rayleigh line has been suppressed) for the different dense AlN samples together with the non-linear square fits to the present peaks. The first interesting observation is that despite the small thickness of the samples, their optical quality and the big aperture of the microscope objective, inherent to the HRmBS technique, the phonon peaks are fairly well-resolved (especially the anti-Stokes side). Moreover, in the spectrum in Fig. 2a four symmetrical Brillouin peaks corresponding to two acoustic modes can be distinguished. In the case of Figs. 3a and 4a only two

symmetrical peaks corresponding to one acoustic mode are visible. Also the better scattering cross section of the amorphous material is evidenced when comparing the three graphs. Due to the sample thickness (markedly below 1 lm), acoustic phonons coupling to the backscattering wave vector and propagating in a direction near the surface normal cannot be obtained. Only phonon coupling to the 2αA wave vector (phonons propagating in the film plane) could be observed. As far as the samples are transparent and several hundreds of nm thick, no information about surface acoustic phonons of the air–AlN interface could be obtained. Therefore, only bulk phonons propagating in the sample plane could be tested. The influence of the substrate on the elastic film properties is one of the aspects that can be of importance if the film thickness is sufficiently low. In our case, the smallest thickness is 325 nm and it is thick enough to make the influence of the substrate on the elastic properties of the film negligible [28]. Moreover, the sound velocity of the bulk shear wave in AlN is very similar to the one of the bulk shear wave in Si (substrate) and therefore its influence is not significant [28].

The Brillouin peaks could correspond to either longitudinal or transverse acoustic phonons propagating within the film plane. The non-linear square fits shown in Figs. 2b, c, 3b, c and 4b, c deliver the related Brillouin frequency shift values that have been used to calculate the sound wave propagation velocity presented in Table 1 and that can be compared to literature values.
From this comparison the observed peaks can be assigned to transverse acoustic modes propagating in the AlN film plane [28, 29]. The spectrum in Fig. 4a corresponds to the amorphous sample and only one transverse acoustic mode can be expected. Similar considerations hold for the polycrystalline with nano-grains sample spectrum in Fig. 3a. In this case the sample behaves as if it was elastically isotropic and only one transverse acoustic mode can be detected. For the nano-columnar sample spectrum in Fig. 2a the elastic relevant symmetry in the film plane is the (a,b) plane of the würtzite-type symmetry and in this case two different transverse acoustic modes can be expected and in fact that is what can actually be observed [30]. In Table 1 the sound propagation velocity values with the corresponding confidence intervals are listed for the three samples. The lowest value corresponds to the polycrystalline nano-granular sample. A similar elastic behaviour was observed previously in the elastic response of PTCa ferroelectric ceramics when passing from amorphous to nano-structured before being fully crystallised [23, 31–33]. In all these cases there is a clear softening of the acoustic longitudinal mode when passing from amorphous to nano-structured. In the case of this work this behaviour is shown by the transverse acoustic mode. The bulk transverse mode values reported in the literature present some dispersion [29], but the experimental values obtained in this work lie clearly lower, which is compatible with the absence of micro-crystallisation in the studied samples.

**CONCLUSIONS**

In conclusion, HRmBS has revealed evident differences in the elastic response of the nano-crystalline/amorphous AlN films. Concerning the influence of the crystallographic structure on the elastic properties, the same pattern as in the case of ferroelectric ceramics is here observed; the acoustic waves present a higher propagation velocity in the case of amorphous structures than in the case of random nanostructured ones. A slightly higher elastic value is observed also in the case of the amorphous AlN sample when compared to the value of the polycrystallised nano-structured one. Nevertheless, the sound propagation velocities of the samples studied here are lower than the generally known values for micro-crystallised samples. This observation can be of interest for future technological applications, especially those using surface acoustic waves.

**TABLES**

| Name of sample | W (Watts) | P (Pa) | Thickness of samples (nm) | Morphology | V (m/s) |
|---|---|---|---|---|---|
| A | 50 | 0.4 | 570 | ncol | 5153 / 6271 |
| B | 75 | 0.3 | 500 | nc | 4911 |
| C | 160 | 0.5 | 325 | a* | 5141 |

Table I. Experimental details of films presented in this study, respectively per column: r.f. power and working pressure during deposition, thickness of AlN deposit, crystalline structure type (referenced in the text) and sound propagation velocity of the transverse modes.

FIGURES and CAPTIONS

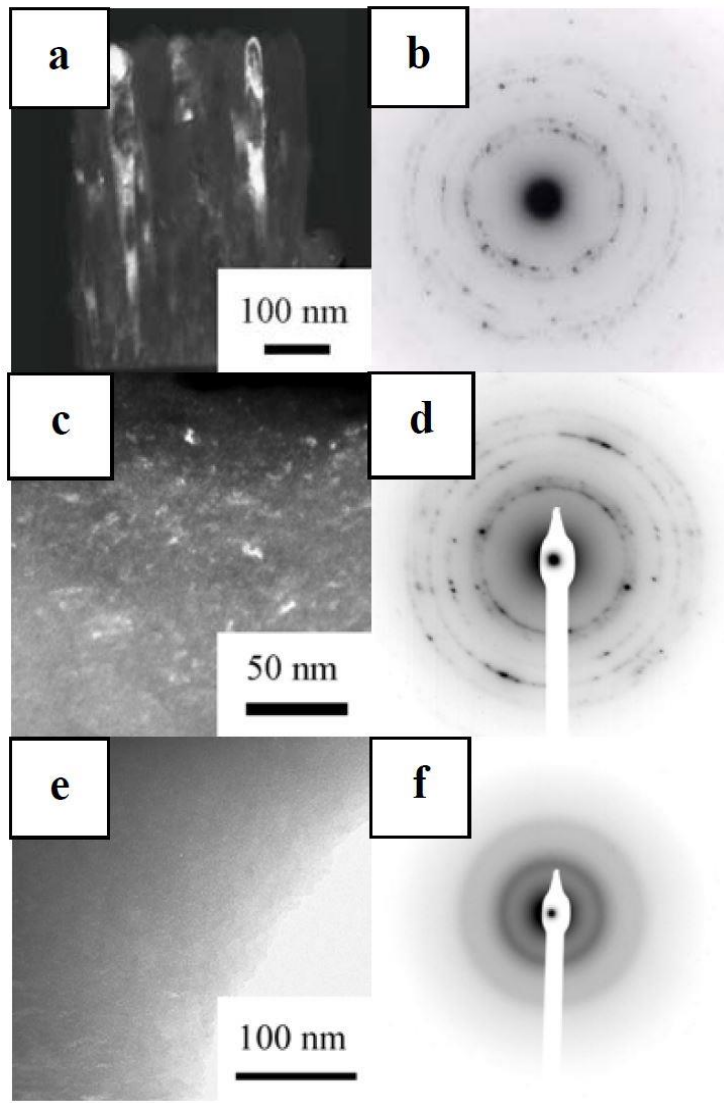

Fig. 1 Microstructure of the three kinds of AlN films studied here. Cross-sectional TEM images and electron diffraction patterns: a Dark field image, b selected area diffraction pattern of the ncol sample, c dark field image, d selected area diffraction pattern of the nc sample, e dark field image, f selected area diffraction pattern of the a* sample (ncol, nc and a* stand for nano-columnar, polycrystallised with small nano-grains and amorphous, respectively)

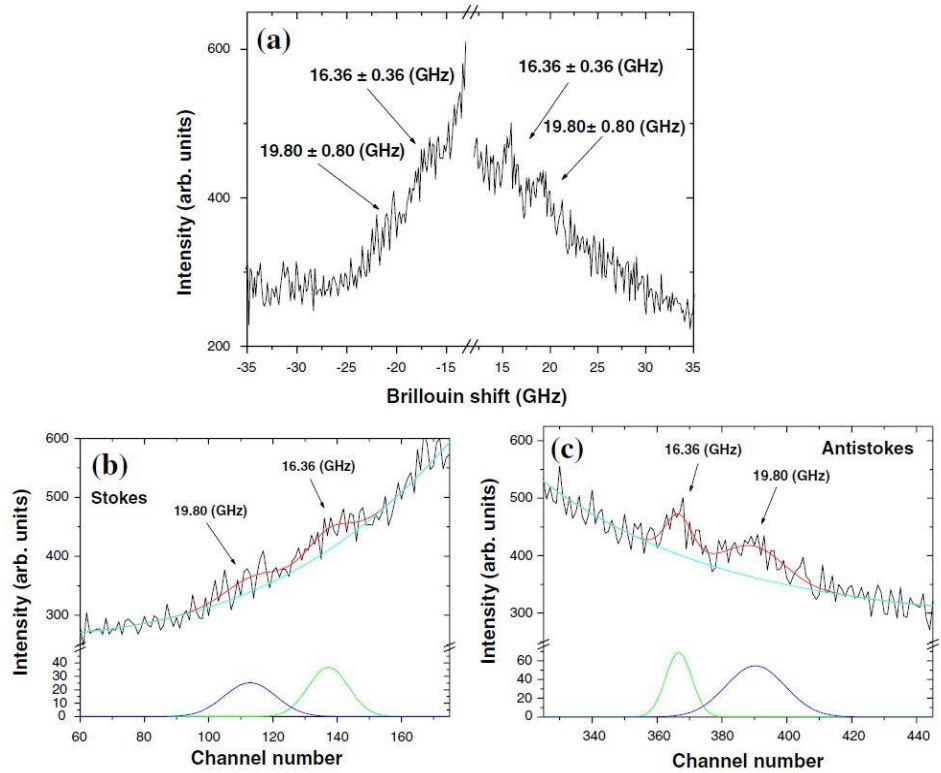

Fig. 2. Fig. 2 a HRmBS spectrum obtained from the nanocolumnar (ncol) sample. The frequency range containing the elastic scattering line or central Rayleigh line has been suppressed and a clear asymmetry in the intensity of both sides is present. Two peaks are clearly seen, especially on the anti-Stokes side of the spectrum. b Stokes side of the spectrum with the result of the non-linear square fit. The *lower side* of the figure presents the two gaussian peaks fitted. c Anti-Stokes side of the spectrum with the result of the non-linear square fit. The *lower side* of the figure presents the two gaussian peaks fitted

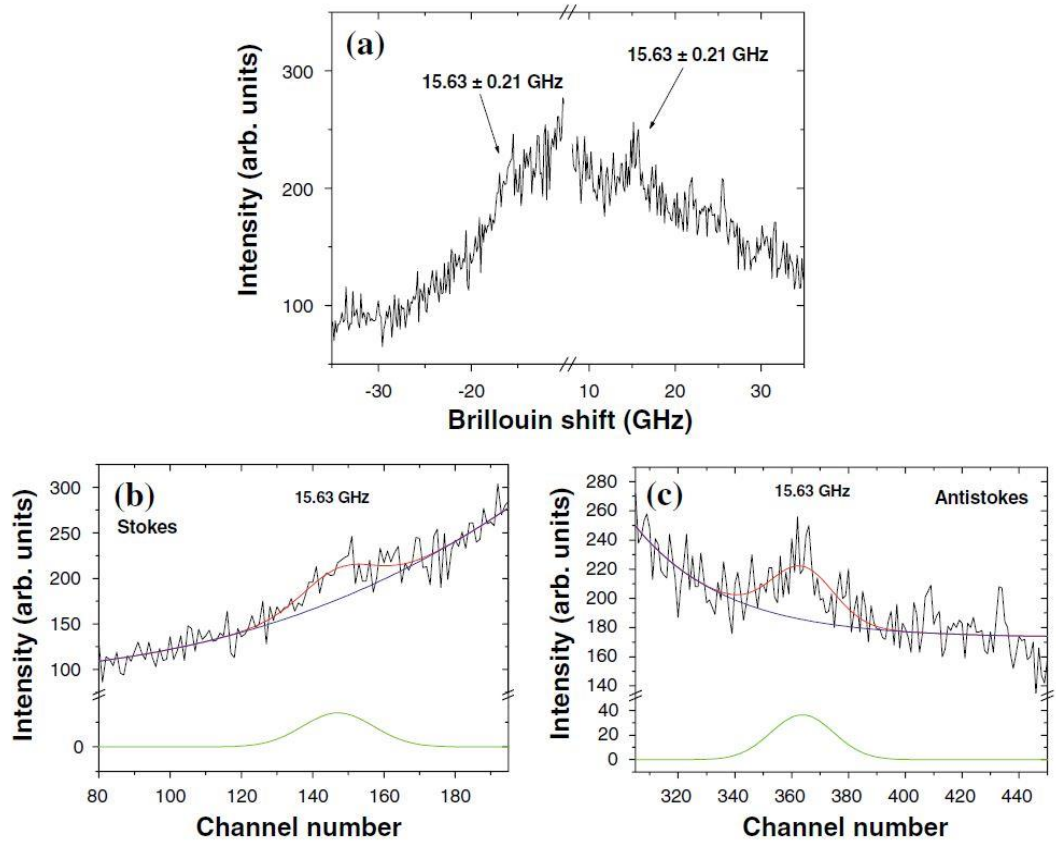

Fig. 3 a HRmBS spectrum obtained from the polycrystallised with small nano-grains (nc) sample. The frequency range containing the elastic scattering line or central Rayleigh line has been suppressed and the clear asymmetry in the intensity of both sides is still present. Only one peak can be seen, especially on the anti-Stokes side of the spectrum. b Stokes side of the spectrum with the result of the non-linear square fit. The *lower side* of the figure presents the gaussian peak fitted. c Anti- Stokes side of the spectrum with the result of the non-linear square fit. The *lower side* of the figure presents the gaussian peak fitted

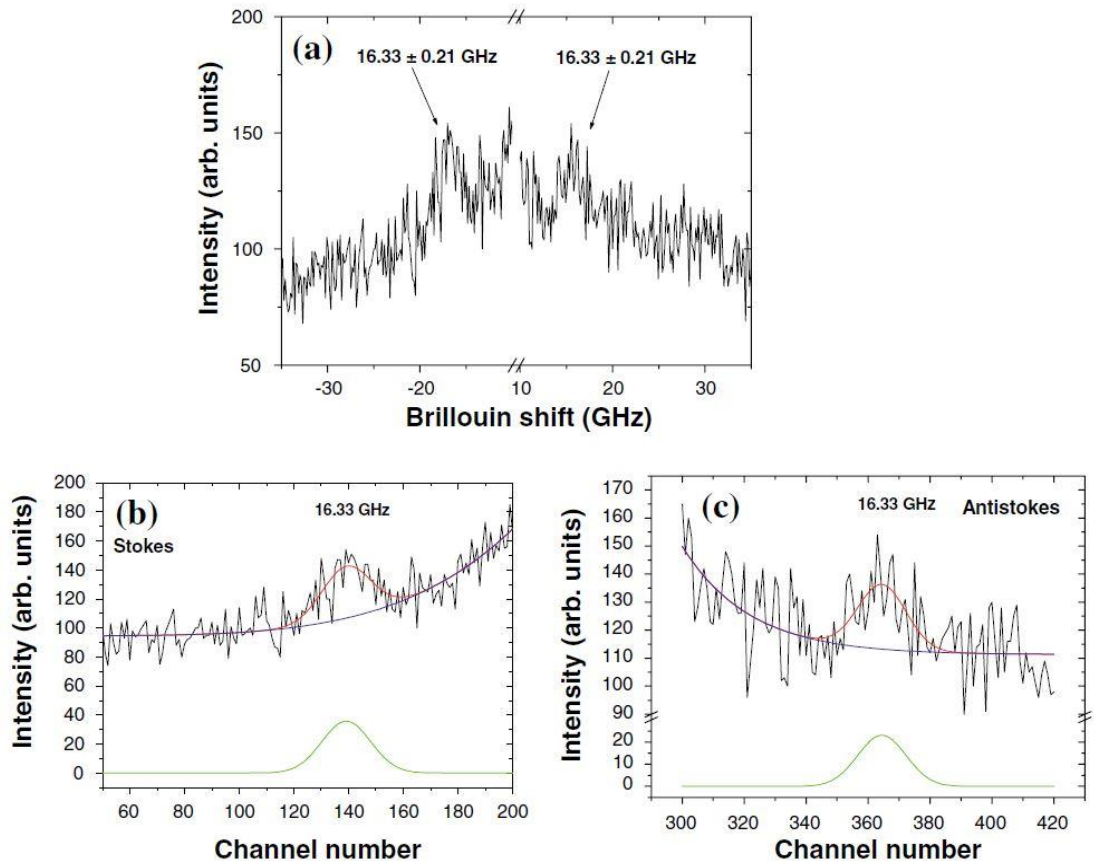

Fig. 4 a HRmBS spectrum obtained from the amorphous (a*) sample. The frequency range containing the elastic scattering line or central Rayleigh line has been suppressed. The asymmetry in the intensity of both sides is less important than in the other samples. Only one peak can be seen, especially on the anti- Stokes side of the spectrum. b Stokes side of the spectrum with the result of the non-linear square fit. The *lower side* of the figure presents the gaussian peak fitted. c Anti-Stokes side of the spectrum with the result of the non-linear square fit. The *lower side* of the figure presents the gaussian peak fitted